\documentclass[symmetry,article,submit,moreauthors,pdftex]{mdpi}
\usepackage{comment}
\usepackage{bbm}
\usepackage{physics}
 \usepackage[force]{feynmp-auto}
\usepackage{scalerel}
\usepackage{diagbox}
\usepackage{datetime} 
\nolinenumbers

\firstpage{1}
\pubvolume{1}
\issuenum{1}
\articlenumber{0}
\pubyear{2023}
\copyrightyear{2023}
\datereceived{}
\dateaccepted{}
\datepublished{}
\hreflink{https://doi.org/
} 
\Title{A novel technique of extracting UCN lifetimes from storage bottle measurements dominated by scattering losses}

\Author{
Prajwal Mohanmurthy, $^{1,2,\orcidA{}}$*, 
Joseph Formaggio $^{2}$,
Daniel J. Salvat $^{3}$,
and Jeff A. Winger $^{4}$
}

\address{%
$^{1}$ \quad Department of Physics, University of Chicago, Chicago, IL 60637, U.S.A\\
$^{2}$ \quad Laboratory for Nuclear Science, Massachusetts Institute of Technology, MA 02139, U.S.A\\
$^{3}$ \quad Center for Exploration of Energy and Matter and Department of Physics, Indiana University, Bloomington, IN 47408, U.S.A\\
$^{4}$ \quad Department of Physics and Astronomy, Mississippi State University, Mississippi State, MS 39762, U.S.A
}

\corres{E-mail: prajwal@mohanmurthy.com}

\abstract{
Neutron lifetime is a critical parameter in the Standard Model. Its measurements using, particularly, the beamline and ultracold neutron storage techniques reveals serious tension. The status of the tension between various measurements have been presented, in light of the insights provided by the $\beta$-decay correlation measurements. When ultracold neutrons are stored in material bottles, they can be lost to various processes, such as $\beta$-decay and up-scattering on material walls. In the past, lifetime extraction by performing a disappearance measurement in material storage bottles, isolated the wall scattering losses by comparing the neutron storage in at least two different storage chambers, with varying volume to surface-area ratios, or characterizing the neutrons lost directly with the help of thermal neutron detectors placed outside the main storage chamber. This technique has been superseded by measurements performed with magneto-gravitational traps, which avoid the wall scattering losses altogether. However, here, we revisit the lifetime measurement in a material storage bottle, dominated by losses from scattering off the walls of the storage chamber. We construct a numerical loss model that includes all the above channels of neutron losses. The neutron energy spectra and its associated uncertainties were, for the first time, well characterized. Input of a center-of-mass offset for the stored ultracold neutron, from the neutron electric dipole moment search, allowed us to further narrow the accepted parameter space. Such models have been used in the extraction of mean time between wall bounces, which is a key parameter for neutron storage disappearance experiments in search of neutron oscillation. A comparison between the loss model and the number of neutrons stored in a single chamber, used for the neutron electric dipole moment search, allowed us to extract a neutron lifetime of $\tau^*_n=879~({+158}/{-78})_{\text{stat.}}~(+230/-114)_{\text{sys.}}~\text{s~~(68.3\% C.I.)}$. Though the uncertainty on this lifetime is not competent with currently available measurements, the highlight of this work is that, we precisely identify the systematic sources of uncertainty that contribute to the neutron lifetime measurements in material storage bottles, namely from the uncertainty in the energy spectra, as well as the storage chamber parameters of Fermi potential and loss per bounce parameter. In doing so, we finally highlight the underestimation of the uncertainties in the previous Monte Carlo simulations of experiments using ultracold neutron storage in material bottles.
}



\begin{document}


\section{Introduction}

Neutrons and protons represent most of the mass of ordinary matter. Of the nucleons, neutrons dominate the mass of terrestrial matter due to an energetically preferred ratio of $n/p \ge 1$ for heavy nuclei due to the electrical charge of the proton \cite{Weizsacker1935-xp}. While the neutron is stable inside the nucleus, free neutrons decay to protons, electrons, and anti-electron neutrino via the charged current mediated $\beta$~decay.

About $1\,$s after the Big Bang, protons and neutrons were frozen out due to their diminishing densities and temperature. The ratio of, $n/p=\exp{-\Delta m/T_{\text{freeze}}}\approx1/6$, was then set, where $\Delta m = m_n - m_p=1.293~$MeV, and $T_{\text{freeze}}=0.782~$MeV \cite{Dubbers1991-hz}. At about $100\,$s after the Big Bang, the universe was cool enough ($T\approx0.1\,$MeV) for neutrons and protons to begin combining to form nuclei \cite{Wietfeldt2011-ay}. By $\sim4\,$minutes after the Big Bang, almost all of the surviving neutrons had combined with protons to form nuclei, chiefly $^4$He due to its high binding energy per nucleon among light nuclei. The ratio of $n/p$ deviates from the ratio at freeze out due to neutron decay, leading to its mass fraction of about $25\%$ \cite{Iocco2009-io}. Details about primordial nucleosynthesis of other light elements may be found in Ref. \cite{Steigman2007-ce,Cyburt2016-od,Izotov2010-zc}. Precise values for the relative primordial abundances of light nuclei is extremely sensitive to the neutron lifetime \cite{Gustavino2016-lm,Serpico2004-ok,Burles1999-qo}.

\newcommand{\quark}[7]{
  \fmfcmd{style_def quarkl#1
      expr p = cdraw subpath (#4) of p shifted (#6);
      cfill (tarrow (p,(xpart(#4)+ypart(#4))*0.48*#5)) shifted (#6);
      if length("#3")=2: label.#3(btex {#2} etex, point ypart(#4) of p shifted (#6)) fi; 
      if length("#3")=3: label.#3(btex {#2} etex, point xpart(#4) of p shifted (#6)) fi; 
    enddef;}
  \fmf{quarkl#1,tension=0}{#7}}
\newcommand{\mylbrace}[2]{\vspace{#2pt}\hspace{2pt}\scaleleftright[\dimexpr6pt+#1\dimexpr0.11pt]{\lbrace}{\rule[\dimexpr2pt-#1\dimexpr0.5pt]{-4pt}{#1pt}}{.}}
\newcommand{\myrbrace}[2]{\vspace{#2pt}\scaleleftright[\dimexpr6pt+#1\dimexpr0.11pt]{.}{\rule[\dimexpr2pt-#1\dimexpr0.5pt]{-4pt}{#1pt}}{\rbrace}\hspace{2pt}}
\unitlength = 0.25mm 
\begin{figure}[h]
\centering
\vspace{3mm}
\begin{fmffile}{feyngraph}
  \begin{fmfgraph*}(120,100)
    \fmfset{arrow_len}{10}
    \fmfstraight
    \fmfleft{i3,i1}
    \fmfright{o3,o2,o1}
    \fmflabel{e$^-(0.511~\text{MeV})$}{o1}
    \fmflabel{$\bar\nu_\text{e}(<0.8~\text{eV})$}{o2}
    \fmf{phantom}{i1,v1,o1}
    \fmf{fermion,tension=0}{o2,v1,o1}
    \fmf{boson,label=W$^-$,label.side=right}{v1,v3}
    \fmfv{l=$\text{n}^0~(939.565~\text{MeV})$\mylbrace{32}{-9},l.d=16,l.a=-160}{i3}
    \fmfv{l=\myrbrace{32}{-9}$\text{p}^+(938.272~\text{MeV})$,l.d=16,l.a=-20}{o3}
    \fmf{phantom}{i3,v3,o3}
    \fmffreeze
    \quark{qai}{d}{lft}{0,1}{0.90}{0,  0}{i3,v3}
    \quark{qbi}{d}{lft}{0,1}{1.00}{0,-10}{i3,v3}
    \quark{qci}{u}{lft}{0,1}{1.10}{0,-20}{i3,v3}
    \quark{qao}{u} {rt}{0,1}{1.10}{0,  0}{v3,o3}
    \quark{qbo}{d} {rt}{0,1}{1.00}{0,-10}{v3,o3}
    \quark{qco}{u} {rt}{0,1}{0.90}{0,-20}{v3,o3}
  \end{fmfgraph*}
\end{fmffile}
\vspace{1cm}
\caption[]{Feynman diagram of the energetically favored $\beta$-decay of the free neutron, as evident from the sum of masses of the products. The masses are from Ref. \cite{Particle_Data_Group2020-an}.}
\label{fig9-0}
\end{figure}
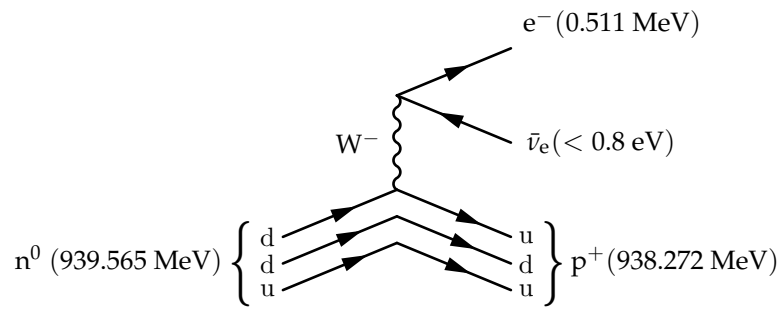

Fermi's four vector $\beta$-decay interaction involving the neutron, proton, electron, and anti-neutrino \cite{Fermi1934-fd} was generalized by Gamov and Teller to involve scalar (S), pseudoscalar (P), tensor (T), axial (A), and vector (V) terms \cite{Gamow1936-lp}. In the modern context of electroweak theory \cite{Salam1964-tq,Weinberg1967-eq,Glashow1971-xw}, the Hamiltonian of the neutron decay shown in Figure~\ref{fig9-0} can be written as a V-A interaction \cite{Feynman1958-yd,Sudarshan1958-hs} by
\begin{equation}
\mathcal{H}=\left[ G_V\bar{p}\gamma_{\mu}n - G_A\bar{p}\gamma_5\gamma_{\mu}n\right]\left[ \bar{e}\gamma_{\mu}\left(1+\gamma_5\right)\nu\right], \label{eq9-1}\
\end{equation}
where $\bar{p}$, $n$, $\bar{e}$, $\nu$ are the four spinors for the states involved in the neutron decay,  $G_V=G_FV_{ud}$, $G_A=G_FV_{ud}\lambda$, $G_F=1.1663787(6)\!\times\!10^{-5}~$GeV$^{-2}$ \cite{Mohr2016-bd} is the Fermi coupling constant, $V_{ud}$ is the first element of the Cabibbo-Kobayashi-Maskawa (CKM) matrix that describes quark mixing \cite{Cabibbo1963-kf,Kobayashi1973-tx}, and $\lambda=G_A/G_V$. One could in principle add the pseudoscalar, scalar, and tensor interactions into Eq.~\ref{eq9-1} but to date there has been no evidence for such interactions. Strong constraints upon scalar and tensor interactions have been placed by studying the $\beta$-decay correlations \cite{Severijns2006-st,Cirigliano2013-do}.

Precise understanding of the nature for the V-A electroweak interaction, along with estimates for the relative premordial abundances of light nuclei, form the core of motivations to measure the neutron lifetime precisely.

\subsection{Previous measurements of Neutron Lifetime}
Neutron decay was first measured using reactor neutrons by Ref. \cite{Snell1950-va} and confirmed by Ref. \cite{Robson1950-oy,Robson1951-kk} using a coincidence measurement of the decay products. Thus far, the most dominant techniques for the neutron lifetime measurements have been: (i) in a beamline where the neutron flux is monitored, and the decay products of electrons, or protons, or both, are counted; and storage of ultracold neutrons (UCN) in (ii)  material bottles, or (iii) magnetic bottles. Neutron lifetime measurements were also made using: cloud chambers where both the decay electrons and protons were tracked \cite{DAngelo1959-hg}; time projection chambers, which could simultaneously measure the neutron flux as well as count the decay electrons and protons \cite{Kossakowski1989-vb}; and using the neutron spectrometer aboard the Messenger spacecraft \cite{Wilson2020-gq}. A detailed overview of the measurements may be found in Refs.~\cite{Wietfeldt2011-ay,Wietfeldt2018-qr}. The experiments discussed here have also been summarized in Table~\ref{tab9-1}.

\begin{table}[h!]
\centering
\caption[]{A summary of the history for measurements of the neutron lifetime, $\tau_n$. Systematic and statistical errors have been combined in quadrature here. If the errors were asymmetric, then the larger of the errors were considered. The measurements marked with: $^*$ were corrected, thus, must be neglected; and $^{\dagger}$ have been considered while averaging measurements in Figure~\ref{fig9-1}.}
\label{tab9-1}
\begin{tabular}{c@{\hspace{1\tabcolsep}}r@{\hspace{0\tabcolsep}}l@{\hspace{0\tabcolsep}}c@{\hspace{0\tabcolsep}}r@{\hspace{0\tabcolsep}}l@{\hspace{0.5\tabcolsep}}c@{\hspace{0.5\tabcolsep}}c@{\hspace{0.5\tabcolsep}}l}
\hline
\hline
Publ.& $(\tau_n$&& $\pm$ & $\sigma_{\tau_n})$ & $s$ & Ref. & Met. & \scriptsize{Notes}\\
\hline
05.1950 & $1730$ & $.$ &$\pm$&$870.$ && \cite{Snell1950-va} & (i) & \scriptsize{$\{p,e\}$ coincidence}\\
05.1950 & $1470$ & $.$ &$\pm$&$690.$ & & \cite{Robson1950-oy} & (i) & \scriptsize{$\vec{E}$ field applied, counted $p$}\\
07.1951 & $1110$ & $.$ &$\pm$&$220.$ && \cite{Robson1951-kk} & (i) & \scriptsize{$\{p,e\}$ coincidence}\\
05.1956 & $1040$ & $.$ &$\pm$&$130.$ && \cite{Spivak1956-qm} & (i) & \scriptsize{Ris{\"o} and Kurch.: $\{p,e\}$ coincidence}\\
02.1959	 & $^* 1013$ & $.$ &$\pm$&$26.$ & & \cite{Sosnovsky1959-wi} & (i) & \scriptsize{Ris{\"o} and Kurch.: $\vec{E}$ field applied, counted $p$}\\
04.1959	& $1100$ & $.$ &$\pm$&$160.$ && \cite{DAngelo1959-hg} & (i) & \scriptsize{Cloud chamber: tracked $\{p,e\}$}\\
04.1972	& $919$ & $.$ &$\pm$&$14.$ & & \cite{Christensen1972-wg} & (i)  & \scriptsize{$4\pi~\beta-$spect., counted $e$}\\
09.1978	 & $^* 877$ & $.$ &$\pm$&$78.$ & &  \cite{Bondarenko1978-ov} & (i) & \scriptsize{Ris{\"o} and Kurch.: upgraded, corrected \cite{Sosnovsky1959-wi}}\\
09.1978	 & $^{\dagger} 891$ & $.$ &$\pm$&$9.$ & & \cite{Spivak1988-le} & (i) & \scriptsize{Ris{\"o} and Kurch.: corrected \cite{Bondarenko1978-ov} for neutron density+proton recoil}\\
05.1980	 & $^* 937$ & $.$ &$\pm$&$18.$ & &  \cite{Byrne1980-ll} & (i) & \scriptsize{Sussex-ILL: Penning trap, counted $p$}\\
03.1988	 & $876$ & $.$ &$\pm$&$21.$ & & \cite{Last1988-ji} & (i) & \scriptsize{PERKEO-I: counted $e$}\\
10.1989	 & $878$ & $.$ &$\pm$&$30.$ & & \cite{Kossakowski1989-vb} & (i) & \scriptsize{He-TPC: counted $e$}\\
07.1990	 & $893$ & $.6$ &$\pm$&$5.$ & $3$  & \cite{Byrne1990-le} & (i) & \scriptsize{Retracted \cite{Byrne1980-ll} due to neutron counting calib.}\\
01.1996	 & $^{\dagger} 889$ & $.2$ &$\pm$&$4.$ & $8$  & \cite{Byrne1996-tl} & (i) & \scriptsize{Sussex-ILL: upgraded}\\
10.2003	 & $886$ & $.8$ &$\pm$&$3.$ & $4$  & \cite{Dewey2003-fk} & (i) & \scriptsize{NIST: quasi-penning trap, counted $p$}\\
05.2005	 & $886$ & $.3$ &$\pm$&$3.$ & $4$  & \cite{Nico2005-ef} & (i) & \scriptsize{NIST: reanalysis}\\
11.2013 & $^{\dagger} 887$ & $.7$ &$\pm$&$2.$ & $2$  & \cite{Yue2013-ed} & (i) & \scriptsize{NIST: upgraded}\\
\hline
02.1980 & $875$ & $.$ &$\pm$&$95.$ & & \cite{Kosvintsev1980-yw} & (ii) & \scriptsize{SM-2: room temp., Al-bottle}\\
11.1986 & $903$ & $.$ &$\pm$&$13.$ &  & \cite{Kosvintsev1986-qq} & (ii) & \scriptsize{SM-2: Al-bottle cooled to $80~$K}\\
08.1989 & $887$ & $.6$ &$\pm$&$3.$ & $0$  & \cite{Mampe1989-ih} & (ii) & \scriptsize{ILL-Mambo: variable geometry piston, fomblin coated}\\
11.1989 & $870$ & $.$ &$\pm$&$8.$ & & \cite{Kharitonov1989-ud} & (ii) & \scriptsize{PNPI: gravitrap, two nested chambers}\\
10.1990 & $^*888$ & $.4$ &$\pm$&$2.$ & $9$  & \cite{Alfimenkov1990-fk} & (ii) & \scriptsize{PNPI: gravitrap, corrected by \cite{Nesvizhevskii1992-nf}}\\
09.1992 & $888$ & $.4$ &$\pm$&$3.$ & $3$  & \cite{Nesvizhevskii1992-nf} & (ii) & \scriptsize{PNPI: gravitrap, reanalysis of \cite{Alfimenkov1990-fk}}\\
01.1993 & $882$ & $.6$ &$\pm$&$2.$ & $7$  & \cite{Mampe1993-cw} & (ii) & \scriptsize{ILL-Kurch.: vertical fomblin coated cyl., thermal neutron det.}\\
06.2000 & $^*885$ & $.40$ &$\pm$&$0.$ & $98$  & \cite{Arzumanov2000-tm} & (ii) & \scriptsize{ILL-Kurch.: horizontal rotating chamber, fomblin grease coated}\\
02.2000 & $^*881$ & $.0$ &$\pm$&$3.$ & $0$  & \cite{Pichlmaier2000-hg} & (ii) & \scriptsize{ILL-Mambo II: Pre-storage chamber added to \cite{Mampe1989-ih}, corrected by \cite{Pichlmaier2001-sr}}\\
01.2005 & $878$ & $.50$ &$\pm$&$0.$ & $76$  & \cite{Serebrov2005-mq} & (ii) & \scriptsize{ILL-PNPI: nested horizontal cyl. and sph. chambers}\\
09.2008 & $^{\dagger}878$ & $.50$ &$\pm$&$0.$ & $76$  & \cite{Serebrov2008-ou} & (ii) & \scriptsize{ILL-PNPI: reanalysis of \cite{Serebrov2005-mq}}\\
10.2010 & $^{\dagger}880$ & $.7$ &$\pm$&$1.$ & $8$  & \cite{Pichlmaier2010-ck} & (ii) & \scriptsize{ILL-Mambo II: cleaned the storage chamber with an absorber}\\
05.2012 & $881$ & $.6$ &$\pm$&$2.$ & $1$  & \cite{Arzumanov2012-eq} & (ii) & \scriptsize{ILL-Kurch.: corrected thermal neutron detector efficiency in \cite{Arzumanov2000-tm}}\\
06.2012 & $^{\dagger}882$ & $.5$ &$\pm$&$2.$ & $1$  & \cite{Steyerl2012-td} & (ii) & \scriptsize{ILL-Mambo: reanalysis of \cite{Mampe1989-ih}, scatt. of UCNs on a liq. wall surface}\\
04.2015 & $^{\dagger}880$ & $.2$ &$\pm$&$1.$ & $2$  & \cite{Arzumanov2015-ty} & (ii) & \scriptsize{ILL-Kurch.: upgraded}\\
05.2018 & $^{\dagger}881$ & $.5$ &$\pm$&$0.$ & $9$  & \cite{Serebrov2018-hs} & (ii) & \scriptsize{ILL-PNPI: large gravitrap, coated with fluorinated fomblin grease}\\
\hline
01.1978 & $918$ & $.$ &$\pm$&$138.$ & & \cite{Kugler1978-qo} & (iii) & \scriptsize{ILL-NESTOR:  magnetic storage ring}\\
01.1985 & $907$ & $.$ &$\pm$&$70.$ & & \cite{Kugler1985-xs} & (iii) & \scriptsize{ILL-NESTOR:  additional statistics \cite{Anton1989-pj}}\\
08.1989 & $877$ & $.$ &$\pm$&$10.$ &  & \cite{Paul1989-gw} & (iii) & \scriptsize{ILL-NESTOR:  upgraded with quad. and sext. magnets}\\
11.1989 & $877$ & $.$ &$\pm$&$10.$ & & \cite{Anton1989-pj} & (iii) & \scriptsize{ILL-NESTOR:  upgraded, sext. torus, colder neutrons}\\
01.2000 & $750$ & $.$ &$\pm$&$330.$ &  & \cite{Huffman2000-ko} & (iii) & \scriptsize{UCNs in superfuild $^4$He, 3D Ioffe-solenoid trap}\\
04.2001 & $660$ & $.$ &$\pm$&$290.$ & & \cite{Brome2001-ro} & (iii) & \scriptsize{Upgrade of \cite{Huffman2000-ko}}\\
01.2006 & $831$ & $.$ &$\pm$&$58.$ & & \cite{Yang2006-en} & (iii) & \scriptsize{Same apparatus as \cite{Huffman2000-ko,Brome2001-ro}}\\
10.2016 & $887$ & $.0$ &$\pm$&$39.$ & & \cite{Leung2016-cx} & (iii) & \scriptsize{LANL: Halback-gravi trap, with cleaning with an absorber}\\
06.2018 & $^{\dagger}878$ & $.3$ &$\pm$&$1.$ & $9$  & \cite{Ezhov2018-iq} & (iii) & \scriptsize{Perm. magnets-solenoid-gravi trap}\\
05.2018 & $877$ & $.7$ &$\pm$&$0.$ & $8$  & \cite{Pattie2018-ss} & (iii) & \scriptsize{LANL: apparatus similar to \cite{Leung2016-cx}}\\
06.2021 & $^{\dagger}877$ & $.75$ &$\pm$&$0.$ & $36$  & \cite{Gonzalez2021-ua} & (iii) & \scriptsize{LANL: similar to \cite{Pattie2018-ss}, additional statistics}\\
\hline
06.2020 & $780$ & $.$ &$\pm$&$92.$ & & \cite{Wilson2020-gq} & & \scriptsize{Messenger spacecraft}\\
\hline
\hline
\end{tabular}
\end{table}

The first dedicated beamline measurements of the neutron lifetime were carried out by the Ris{\"o} and Kurchatov group \cite{Spivak1956-qm,Sosnovsky1959-wi,Bondarenko1978-ov}, where the later two upgrades extracted the decay protons with an electric field before counting them. Similarly, $\beta$-decay spectroscopy measurements where the decay electrons were counted also lead to measurements of neutron lifetime \cite{Christensen1972-wg}. The first neutron lifetime crisis emerged when the penning trap measurements at the Institut Laue-Langevin (ILL) \cite{Byrne1980-ll,Byrne1990-le} disagreed with the measurements in Ref.~\cite{Christensen1972-wg}. However the first crisis was resolved after a number of measurements corrected or updated the neutron lifetime \cite{Spivak1988-le,Byrne1996-tl}, which ultimately confirmed the lower value around $890\,$s. The PERKEO collaboration, known for their measurements of the $\beta$~asymmetry in neutron decay, also published a value for the neutron lifetime in Ref.~\cite{Last1988-ji}. The most recent measurements using the beamline technique come from efforts at the National Institute of Standards and Technology (NIST) \cite{Dewey2003-fk,Nico2005-ef,Yue2013-ed}, which used a quasi-penning trap and benefited from the very large cold neutron flux at NIST, and has culminated in the measurement of lifetime precise to $2.2\,$s.

The first neutron lifetime measurements using UCNs were made at the SM-2 reactor using an aluminum storage bottle at room temperature \cite{Kosvintsev1980-yw}, and the subsequent upgraded measurements spanned nearly a decade \cite{Kosvintsev1986-qq,Morozov1989-rh}. The neutron counting systematic in these measurements is dominated by the up-scattering of UCN on the walls of the storage bottle, and therefore many efforts have employed various techniques to better characterize the scattering losses. The Mambo collaboration at ILL performed many neutron lifetime measurements using a bottle with variable geometry in order to characterize the UCN scattering losses \cite{Mampe1989-ih,Pichlmaier2000-hg,Pichlmaier2010-ck,Steyerl2012-td}. The group at Petersburg Nuclear Physics Institute (PNPI) used a gravitrap, where neutrons are vertically trapped in a gravitational potential well to measure the lifetime by extrapolating the UCN velocity to zero \cite{Kharitonov1989-ud,Alfimenkov1990-fk,Nesvizhevskii1992-nf,Serebrov2018-hs}. The measurements by the ILL-Kurchatov group characterized the UCN scattering losses by placing a thermal neutron detector outside their trap \cite{Mampe1993-cw,Arzumanov2000-tm,Arzumanov2012-eq,Arzumanov2015-ty}, while an ILL-PNPI effort used two nested storage chambers of differing geometry \cite{Serebrov2005-mq,Serebrov2008-ou}, and ultimately brought down the uncertainty associated with the lifetime to $0.8\,$s.

The last category of measurements use the technique of UCN storage in magnetic bottles such that the wall scattering loss channels are avoided. The first such measurement was performed using the NESTOR magnetic storage ring at ILL \cite{Kugler1978-qo} and the subsequent upgrades \cite{Kugler1985-xs,Paul1989-gw,Anton1989-pj} made the precision of these measurements competitive when compared with the then best measurements using the previous two techniques. An independent effort by the ILL-PNPI group used a cylindrical storage vessel with permanent magnets trapping the UCNs in the transverse direction and solenoid magnetic along with gravity trapping them in the vertical (longitudinal) direction \cite{Ezhov2009-pk,Ezhov2018-iq}. A notable effort using an Ioffe-solenoid trap in which the UCNs are stored in superfuild $^4$He has also reported lifetime measurements \cite{Huffman2000-ko,Brome2001-ro,Yang2006-en}. The most recent effort in this category uses a Halbach magnetic array to trap the neutrons in the transverse direction and on the bottom, while trapping them vertically within a gravitational potential well \cite{Leung2016-cx,Pattie2018-ss,Gonzalez2021-ua}, achieving an impressive precision of $0.36\,$s.

\subsection{Tension between beam-line and UCN lifetime measurements}

A second neutron lifetime crisis emerged with one of the most precise measurements for the neutron lifetime of $\tau_n=(878.50\pm0.76)\,$s in 2004 \cite{Serebrov2005-mq}.  This value was in conflict with the then global average lifetime of $\tau_n=(885.7\pm0.8)\,$s. The 2004 measurement by Ref.~\cite{Serebrov2005-mq} was not the first with a precision under $1\,$s.  Ref.~\cite{Arzumanov2000-tm} reported the first measurement with a lifetime precise to under $1\,$s; however, the later measurement was subsequently corrected by Ref.~\cite{Arzumanov2012-eq} to a precision of $2.1\,$s due to corrections in the thermal neutron detector efficiency. Measurement by Ref.~\cite{Serebrov2005-mq} has since been confirmed by various experiments in Refs.~\cite{Ezhov2018-iq,Pattie2018-ss,Gonzalez2021-ua} and by Refs.~\cite{Pichlmaier2010-ck,Steyerl2012-td,Arzumanov2015-ty} within $2$ standard deviations. However, the latest experiment using UCNs and a new large gravitrap \cite{Serebrov2018-hs} by the same group only agrees with their measurement in 2004 within $3$ standard deviations.

\begin{figure}[h]
\includegraphics[width=\columnwidth]{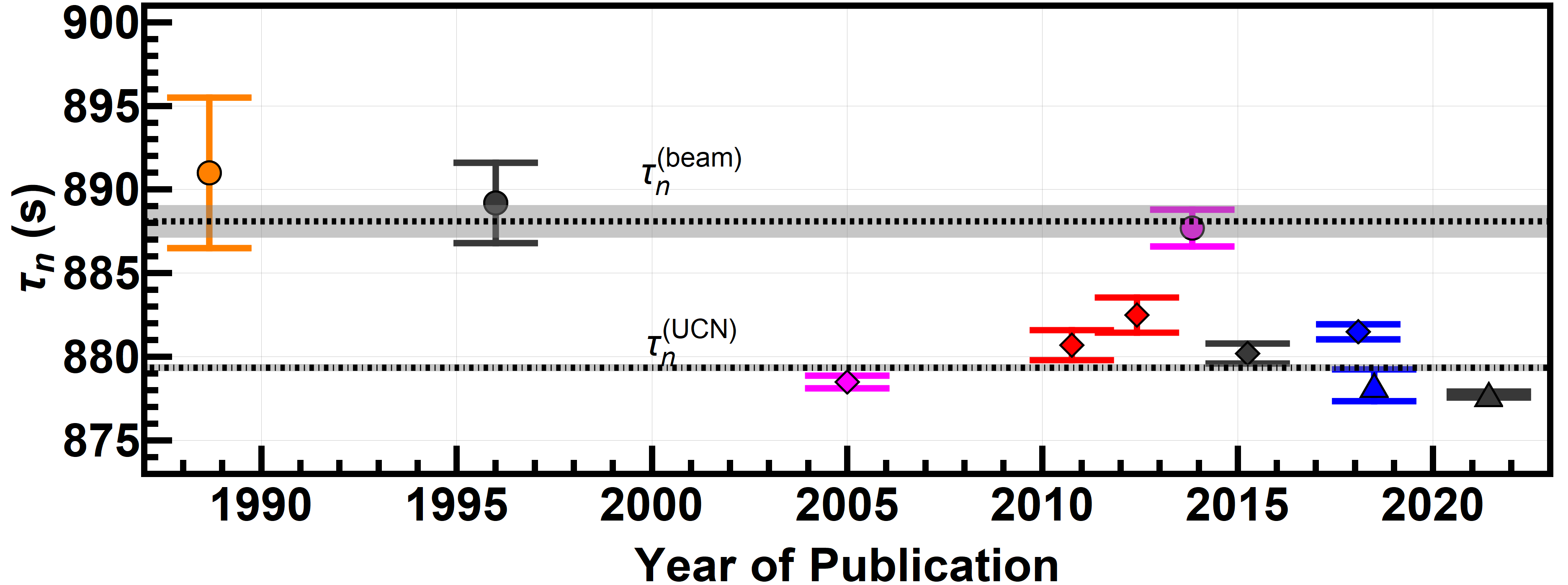}
\caption[Measurements of neutron lifetime using the two techniques involving beamline (shown as circles), and UCN storage, both in material (shown as diamonds) and in magnetic (shown as triangles) bottles.]{Measurements of neutron lifetime using the two techniques involving beamline (shown as circles), and UCN storage, both in material (shown as diamonds) and in magnetic (shown as triangles) bottles. The dashed (dot-dashed) line and the associated gray shaded area represent the respective weighted averages in Eq.~\ref{eq9-4} (\ref{eq9-3}) and their uncertainty for the measurements using the beamline (UCN storage) technique.}
\label{fig9-1}
\end{figure}

The measurements for the neutron lifetime in Refs.~\cite{Serebrov2008-ou,Pichlmaier2010-ck,Steyerl2012-td,Arzumanov2015-ty,Serebrov2018-hs} are the most up-to-date iterations of the measurements using their respective apparatus and the subsequent upgrades, where the technique employed was UCN storage in material bottles. Similarly, measurements in Refs.~\cite{Ezhov2018-iq,Gonzalez2021-ua} were made using the technique of UCN storage in magnetic bottles. The weighted average of these seven measurements that used UCNs is
\begin{equation}
\tau^{(\text{UCN})}_n = (878.56 \pm 0.29)~\text{s}~,~\chi^2/ndf=2.78 \label{eq9-3}
\end{equation}
For above seven measurements, only the measurement in Ref.~\cite{Serebrov2018-hs} deviates from the average in Eq.~\ref{eq9-3} by more than $3$ standard deviations. If the measurement in Ref.~\cite{Serebrov2018-hs} is dropped from the weighted average, then the neutron lifetime using UCNs drops to $(878.22\pm0.31)~$s, $(\chi^2/ndf=1.60)$. But without a compelling argument to support leaving out the measurement in Ref.~\cite{Serebrov2018-hs}, we have retained the value. Measurements in Refs.~\cite{Byrne1996-tl,Kossakowski1989-vb,Yue2013-ed} which used the technique of counting the neutron decay products in a beamline all agree within one standard deviation. Their weighted average is
\begin{equation}
\tau^{(\text{beam})}_n = (888.1 \pm 2.0)~\text{s}~,~\chi^2/ndf=0.12 \label{eq9-4}
\end{equation}
Certain measurements whose uncertainties are so large that they change the respective weighted averages by less than $1\%$, \emph{e.g.} Refs.~\cite{Last1988-ji,Kossakowski1989-vb}, have been left out of the above averages despite being the most up-to-date measurement from the particular apparatus. The seven UCN measurements and the three beamline measurements considered above, along with their respective weighted averages in Eqs.~\ref{eq9-3} and \ref{eq9-4} are shown in Figure~\ref{fig9-1}. Since the uncertainty associated with the beamline measurements is over six times larger than that associated with the UCN measurements, and the UCN measurements being more recent, usually, the weighted average from the UCN measurements is preferred.

The averages in Eqs.~\ref{eq9-3} and \ref{eq9-4} disagree by about $4.5$ standard deviations, and are thus incompatible. Without a way to reconcile the tension between the two families of measurements, the second neutron lifetime crisis, that emerged with the measurement of Ref.~\cite{Serebrov2005-mq}, remains to this day. A meta-analysis of this disagreement can be found in Ref.~\cite{Rajan2020-mj}. Various explanations have been proposed to reconcile the two families of measurements including: dark matter \cite{Fornal2018-sb}; neutron to mirror-neutron oscillations \cite{Berezhiani2019-ot}; and other exotic neutron decay channels \cite{Berezhiani2019-cq,Czarnecki2018-ck}. However, it has also been pointed out by Ref.~\cite{Dubbers2019-je} that exotic decay channels cannot be the reason for this disagreement.

\subsection{Inputs into the neutron lifetime from measurements of $\beta$-decay correlations}

Beta decay of free neutrons provides a way to precisely probe the weak interaction of the standard model, particularly the ratio between the axial-to-vector weak-coupling constants, $\lambda$, and the CKM-matrix element, $V_{ud}$. Such studies are done by measuring the various correlation coefficients between 4-momenta for the decay products of protons, electrons and neutrinos. The rate of $\beta$-decay is linked to these correlation coefficients as \cite{Jackson1957-le}
\begin{eqnarray}
\frac{dw}{dE_e} & \propto & \frac{2\pi}{\hbar} G_F^2 V^2_{ud} \rho(E_e) \cdot \left( 1 + 3 |\lambda|^2\right) \cdot \nonumber \\
&& \left\{ 1 + a \frac{\vec{p}_e \cdot \vec{p}_{\nu}}{E_e E_{\nu}} + b\frac{m_e}{E_e} + \vec{\sigma}_n\cdot \left( A\frac{\vec{p}_e}{E_e} + B\frac{\vec{p}_{\nu}}{E_{\nu}} + C\frac{\vec{p}_{p}}{E_{p}} + N\vec{\sigma}_e\right)\right\}, \label{eq9-5}
\end{eqnarray}
where $w$ is the $\beta$-decay rate, $\rho(E_e)$ is the energy distribution of the electron, $\vec{\sigma}_n$ is the polarization state of the initial neutron, $\vec{\sigma}_e$ is the electron polarization state, $A = -2(|\lambda|^2+Re(\lambda))/(1+3|\lambda|^2)$ is the $\beta$-electron asymmetry, $a = (1 - |\lambda|^2)/(1 + 3|\lambda|^2)$ is the neutrino-electron correlation, $B = 2(|\lambda|^2-|\lambda|)/(1 + 3|\lambda|^2)$ is the neutrino asymmetry, $C = 0.27484\times(4|\lambda|)/(1 + 3|\lambda|^2)$ is the proton asymmetry, and $N = A\gamma$, and $\gamma$ is the Lorentz boost. Additional terms can be added to Eq.~\ref{eq9-3}, such as accounting for $T$ symmetry violation and terms involving more than two vectors \cite{Severijns2006-st}, but such terms are beyond the scope of this work. Recent progress in measuring the $\beta$-decay correlation coefficients has been summarized in Ref.~\cite{Gonzalez-Alonso2019-rf}.

\begin{table}[h!]
\centering
\caption[]{A summary of the history for measurements of $\lambda$. The measurements marked with: $^*$ were corrected, thus, must be neglected; and $^{\dagger}$ have been considered for measurements shown in Figure~\ref{fig9-2}.}
\label{tab9-2}
\begin{tabular}{c@{\hspace{1\tabcolsep}}r@{\hspace{0\tabcolsep}}l@{\hspace{0\tabcolsep}}c@{\hspace{0\tabcolsep}}r@{\hspace{0\tabcolsep}}l@{\hspace{1\tabcolsep}}c@{\hspace{1\tabcolsep}}l}
\hline
\hline
Publ.& $(\lambda$&& $\pm$ & $\sigma$ & $_{\lambda})$ & Ref. & \scriptsize{Notes}\\
\hline
03.1986 & $^{\dagger}1$ & $.262$ &$\pm$&$0$ & $.005$ & \cite{Bopp1986-wa} & \scriptsize{PERKEO-I: measured $A$}\\
10.1997 & $^{\dagger}1$ & $.2594$ &$\pm$&$0$ & $.0038$ & \cite{Yerozolimsky1997-vc} & \scriptsize{PNPI-Kurch.: measured $A$}\\
10.1997	& $^{\dagger}1$ & $.266$ &$\pm$&$0$ & $.004$ & \cite{Liaud1997-vb} & \scriptsize{ILL-TPC: measured $A$}\\
09.2001 & $^{\dagger}1$ & $.2686$ &$\pm$&$0$ & $.0046$ & \cite{Mostovoi2001-sq} & \scriptsize{PNPI-Kurch.-ILL: measured $A,B$}\\
04.2008 & $^{\dagger}1$ & $.275$ &$\pm$&$0$ & $.016$ & \cite{Schumann2008-zm} & \scriptsize{PERKEO-II: measured $C$}\\
04.2013 & $^{\dagger}1$ & $.2761$ &$\pm$&$0$ & $.0017$ & \cite{Mund2013-gh} & \scriptsize{PERKEO-II: measured $A$}\\
07.2017 & $^{\dagger}1$ & $.284$ &$\pm$&$0$ & $.014$ & \cite{Darius2017-lk} & \scriptsize{aCORN: measured $a$}\\
03.2018 & $^{\dagger}1$ & $.2772$ &$\pm$&$0$ & $.0020$ & \cite{UCNA_Collaboration2018-br} & \scriptsize{UCNA: measured $A$}\\
06.2019 & $^{\dagger}1$ & $.27641$ &$\pm$&$0$ & $.00056$ & \cite{Markisch2019-bk} & \scriptsize{PERKEO-III: measured $A$}\\
05.2020 & $^{\dagger}1$ & $.2677$ &$\pm$&$0$ & $.0028$ & \cite{Beck2020-cp} & \scriptsize{aSPECT: measured $a$}\\
\hline
01.1975 & $1$ & $.258$ &$\pm$&$0$ & $.015$ & \cite{Krohn1975-wm} & \scriptsize{ANL-CP5: measured $A$}\\
02.1975 & $1$ & $.250$ &$\pm$&$0$ & $.036$ & \cite{Dobrozemsky1975-hm} & \scriptsize{Vienna-ASTRA: measured $a$, updated by \cite{Stratowa1978-ua}}\\
06.1976 & $1$ & $.263$ &$\pm$&$0$ & $.015$ & \cite{Erozolimskii1976-jn} & \scriptsize{Kurch.: measured $A$, updated by \cite{Erozolimsky1979-fd}}\\
12.1978 & $1$ & $.259$ &$\pm$&$0$ & $.017$ & \cite{Stratowa1978-ua} & \scriptsize{Vienna-ASTRA: measured $a$}\\
01.1979 & $1$ & $.261$ &$\pm$&$0$ & $.012$ & \cite{Erozolimsky1979-fd} & \scriptsize{Kurch.: measured $A$}\\
02.1983 & $1$ & $.226$ &$\pm$&$0$ & $.042$ & \cite{Mostovoi1983-vh} & \scriptsize{PNPI-Kurch.-ILL: measured $A,B$, updated by \cite{Mostovoi2001-sq}}\\
07.1991 & $^*1$ & $.2544$ &$\pm$&$0$ & $.0036$ & \cite{Erozolimskii1991-pe} & \scriptsize{LNPI-Kurch.: measured $A$, corrected by \cite{Yerozolimsky1997-vc}}\\
04.1995 & $1$ & $.266$ &$\pm$&$0$ & $.004$ & \cite{Schreckenbach1995-fv} & \scriptsize{ILL-TPC: measured $A$}\\
09.1997 & $1$ & $.274$ &$\pm$&$0$ & $.003$ & \cite{Abele1997-co} & \scriptsize{Perkeo-II: measured $A$}\\
05.2002 & $1$ & $.2739$ &$\pm$&$0$ & $.0019$ & \cite{Abele2002-vb} & \scriptsize{Perkeo-II: measured $A$}\\
11.2012 & $1$ & $.2759$ &$\pm$&$0$ & $.0045$ & \cite{UCNA_Collaboration2012-eb} & \scriptsize{UCNA: measured $A$ \cite{Liu2010-oq}}\\
03.2013 & $1$ & $.2755$ &$\pm$&$0$ & $.0030$ & \cite{UCNA_Collaboration2013-ez} & \scriptsize{UCNA: measured $A$}\\
\hline
\hline
\end{tabular}
\end{table}

Taking into account the radiative corrections to the neutron $\beta$~decay links the neutron lifetime, $\tau_n$, to the values of $V_{ud}$ and the relative strength of the axial-to-vector coupling constant, $\lambda$, \cite{Marciano2006-cy,Czarnecki2018-ck} through
\begin{equation}
\tau_n=\frac{4908.7\pm1.9~\text{s}}{|V_{ud}|^2\left(1+3|\lambda|^2\right)}. \label{eq9-2}
\end{equation}
The neutron lifetime therefore gives direct input into the fundamental parameters of the standard model like $V_{ud}$ and $\lambda$. Measurement of the $\beta$-decay correlation coefficients also provide a value for $\lambda$, as can be seen in Eq.~\ref{eq9-5} and the subsequent descriptions of the correlation coefficients. Therefore, we can use the value of $\lambda$ obtained from the measurement of the $\beta$-decay correlation coefficients along with independent measurements of $V_{ud}$ to shed some additional light on the issue of the disagreeing neutron lifetime measurements.

\begin{figure}[h]
\includegraphics[width=\columnwidth]{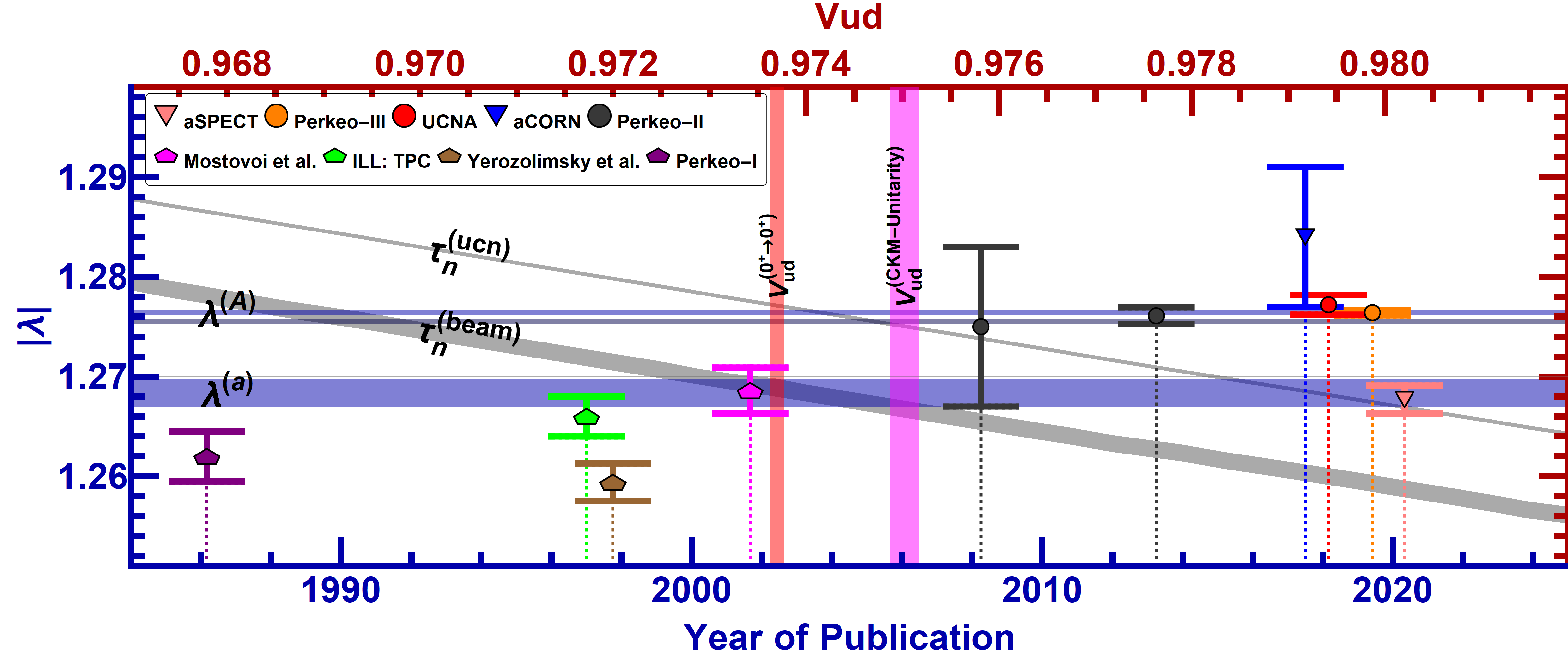}
\caption[]{Figure showing the values of $V_{ud}$ and $\lambda$ extracted from the various neutron lifetime and $\beta$-decay correlation measurements. The diagonal gray bars use the lifetime measurements. The horizontal blue bars are best fit values of $\lambda$ pre-2002 and post-2002. Vertical red bar is the measure of $V_{ud}$ from global analysis of $\beta$~decay between $0^+\rightarrow0^+$ states.}
\label{fig9-2}
\end{figure}

The values of $\lambda$ obtained from $\beta$-decay correlation measurements have been summarized in Table~\ref{tab9-2}, along with the ten most up-to-date measurements, each from a unique apparatus. A summary of measurements before 1972 can be found in Ref.~\cite{Kropf1974-ex}. A weighted average of the most up-to-date measurements is
\begin{eqnarray}
\langle \lambda \rangle = (1.27550\pm0.00049)~,~\chi^2/ndf=4.28. \label{eq9-6}
\end{eqnarray}
However, as noted by Ref.~\cite{UCNA_Collaboration2018-br}, measurements from before 2002 had larger than $10\%$ systematic corrections. Neglecting the four measurements from before 2002 and Ref.~\cite{Schumann2008-zm}, which obtains $\lambda$ through the correlation coefficient $C$, the weighted averages of $\lambda$ from the three experiments that measure $A$ and from the two experiments that measured $a$ are
\begin{eqnarray}
\langle \lambda \rangle^{(A)} &=& (1.27643\pm0.00051)~,~\chi^2/ndf=0.06, \label{eq9-7}\\
\langle \lambda \rangle^{(a)} &=& (1.2683\pm0.0027)~,~\chi^2/ndf=0.62. \label{eq9-8}
\end{eqnarray}
The above averages of $\lambda$ differ by $3$ standard deviations and are thus incompatible. Since the average of $\lambda$ obtained from experiments measuring $A$ is over $5$ times more precise than that from experiments measuring $a$, the global weighted average in Eq.~\ref{eq9-6} is heavily weighted by the former measurements. The weighted averages in Eqs.~\ref{eq9-7} and \ref{eq9-8}, the five measurements used to calculate them from Table~\ref{tab9-2}, along with Eq.~\ref{eq9-2} for the two values of neutron lifetime in Eqs.~\ref{eq9-3} and \ref{eq9-4} have been presented in Figure~\ref{fig9-2}.

Figure~\ref{fig9-2} also displays the values of $V_{ud}$ obtained from two independent methods: from $0^+ \rightarrow 0^+$ super-allowed decays \cite{Particle_Data_Group2020-an,Hardy2015-ex}, and via CKM-unitarity ($V^2_{ud}+V^2_{us}+V^2_{ub}=1$) by using the PDG recommended values of $V_{us}=0.2221\pm0.0013$ \cite{Particle_Data_Group2020-an} (average of both inclusive and exclusive strangeness changing $\tau$ decays) and $V_{ub}=(3.82\pm0.24)\times10^{-3}$ \cite{Particle_Data_Group2020-an} (average of both inclusive and exclusive $\bar{B}$ decays).  These values are
\begin{eqnarray}
V_{ud}^{(0^+ \rightarrow 0^+)} & = & (0.97370\pm0.00014), \qquad\mathrm{and}\label{eq9-9}\\
V_{ud}^{\text{(CKM-Unitarity)}} = \sqrt{1-V^2_{us}-V^2_{ub}} & = & (0.97502\pm0.00030). \label{eq9-10}
\end{eqnarray}
The values of $V_{ud}$ also differ by over $4$ standard deviations ($4.4\sigma$), and are thus incompatible. The evolving status of CKM-unitarity maybe found in Refs.~\cite{Towner2010-za,Hardy2013-po,Belfatto2020-jr}.

Looking at Figure~\ref{fig9-2}, a case can be made for either of the two neutron lifetimes. The UCN neutron lifetime taken with the value of $V_{ud}$ obtained from $0^+ \rightarrow 0^+$ super-allowed decays, and the value of $\lambda$ from experiments measuring $A$ is consistent in the $95\%$ ($\sim2\sigma$) confidence region. On the other hand, the beam neutron lifetime taken along with the value of $V_{ud}$ obtained from either $0^+ \rightarrow 0^+$ super-allowed decays as well as CKM-unitarity, and the value of $\lambda$ from experiments measuring $a$ is consistent in the $68.3\%$  ($1\sigma$) confidence region. Also, the UCN neutron lifetime is consistent within the $68.3\%$  ($1\sigma$) confidence region, when taken together with the value of $V_{ud}$ obtained from CKM-unitarity, and the global value of $\lambda$ in Eq.~\ref{eq9-6}.

\section{Apparatus}

We used data from Ref.~\cite{Abel2019-rk}, which utilized the ultra-cold neutron (UCN) source \cite{Anghel2009-iu,Lauss2021-ge} at the Paul Scherrer Institute (PSI) and the neutron electric dipole moment (nEDM) apparatus \cite{Baker2014-hd,Abel2019-cv}. In Ref.~\cite{Abel2019-rk}, unpolarized UCNs were employed, and the UCNs were stored in a cylindrical chamber of height, $h = 12.000(1)\,$cm, and radius, $R = 23.500(1)\,$cm. Also in Ref.~\cite{Abel2019-rk}: after a period of storage, the remaining UCNs were counted by a detector system, and the electric and magnetic fields were turned off. The residual magnetic field was constrained to less than $0.36~\upmu$T \cite{Abel2019-rk}.

Further information regarding the $^{199}$Hg and $^{133}$Cs magnetometers used in the nEDM experiment at PSI can be found in {refs.} \cite{Ban2018-ur,Abel2020-ig}, and the neutron detector system is described in {refs.}~\cite{Afach2015-lz,Ban2016-xw}. 
Information regarding the magnetic field and its compensation may be found in {refs.}~\cite{Abel2019-oh,Abel2022-tl,Afach2014-ji}, and details of the settings used, while data may be found in Ref.~\cite{Abel2019-cv}.

\section{Analysis: Fitting the loss model to the stored UCNs}

\begin{figure}[p]
\includegraphics[width=\columnwidth]{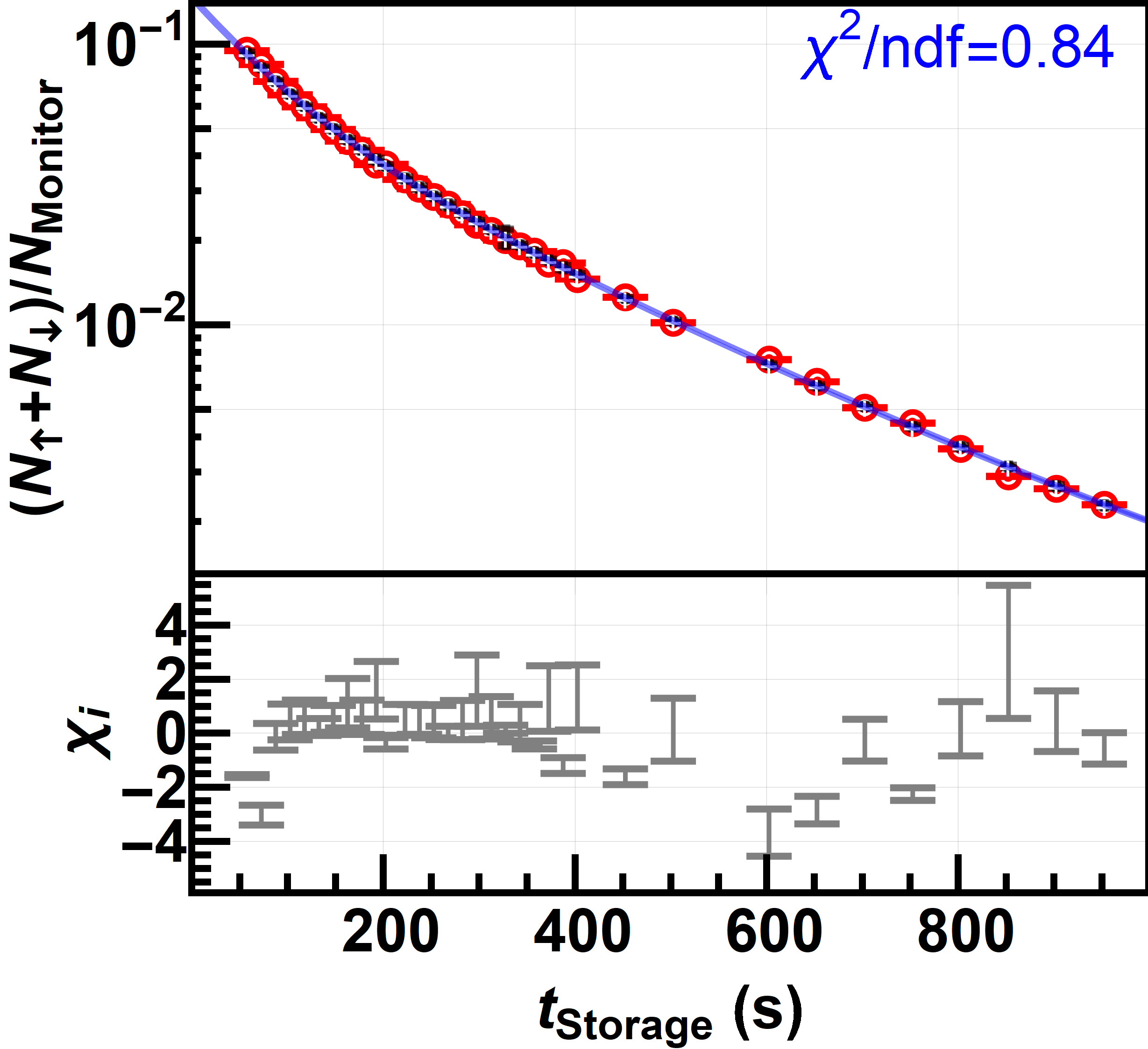}
\caption[]{(Top) The decay curve showing the monitor-normalized neutrons counted as functions of storage time, $t_s$. The black (cross) data points are the experimental values measured using the nEDM storage chamber in Ref.~\cite{Abel2019-rk}. Each red (circle) data points (Top) represents the best value of $n'(t_s)$ in Equation~(\ref{eq9-5-5-7-1-c}), for a particular storage time, obtained using the numerical loss model, as explained in the text. The blue curve (and the associated reduced $\chi^2$) corresponds to the double exponential fit yielding the lifetimes in Equation~(\ref{eq9-12}). (Bottom) Plot showing the range of residuals obtained by comparing each of the experimental values (black crosses) with a set of numerical loss model values of $n'(t_s)$ in Equation~(\ref{eq9-5-5-7-1-c}) (the best value of which is shown as red circles).}
\label{fig9-3}
\end{figure}

The analysis presented here uses the storage data reported in Ref.~\cite{Abel2019-rk} (blue curve in Figure 7) which is represented here in Figure~\ref{fig9-3} (Top), in conjunction with the Kassiopeia simulation toolkit \cite{kassiopeia}, specifically the version adapted for use in UCN simulations presented. In Ref.~\cite{Abel2019-rk}, unpolarized UCNs were stored for a fixed storage time, $t_s$, before being counted by the detector systems. Also the total number of neutrons counted after a storage time was normalized using the source monitor, to account for the varying neutron fluxes from the source \cite{Abel2019-rk}. The normalized neutron count, as a function of storage time, is usually referred to as the \emph{storage curve}. The two fitted time constants are
\begin{eqnarray}
T_1 &=& (80.6062~~\pm0.0002)~\text{s}, \qquad\mathrm{and}\label{eq9-11}\\
T_2 &=& (271.57161\pm0.00003)~\text{s}. \label{eq9-12}
\end{eqnarray}
The short time constant roughly represents the loss of higher energy UCNs quickly, and the longer time constant roughly represents the scale of time over which neutrons are lost in the chamber due to various effects such as neutron decay and losses due to scattering off the walls. In order to isolate the neutron decay lifetime, a model for the other loss channels would need to be developed. This section describes the loss model for the stored UCNs, based on losses when the UCNs scatter inside the storage chamber by first characterizing the energy spectra of the stored UCNs and then also constraining the model with the center-of-mass offset obtained in the nEDM analysis, before finally presenting the possible values of neutron lifetime that would be consistent with the loss model.

An ensemble of UCNs stored in the storage chamber has an energy distribution determined by the spectrum from the UCN source, the time of storage, $t_s$, and the storage chamber wall surface quality. The storage lifetime resulting from different losses can be written as
\begin{eqnarray}
\frac{1}{\tau_s} &=& \frac{1}{\tau_{n}} + \frac{1}{\tau_{\mu}} + 
\ldots,\label{eq1-4-1-2}
\end{eqnarray}
where $\tau_n$ is the free neutron decay lifetime, and $\tau_{\mu}$ corresponds to the loss channel due to scattering off the walls. Further losses can be due to absorption or leakage of neutrons, but these sources are negligible in the context of the sensitivity of this effort. The parameter $\mu$ is linked to the energy independent loss coefficient: $\eta_s = W/V_F$, and is given as an averaging over all angles of incidence~\cite{Golub1991-ap} by
\begin{eqnarray}
\mu(E_{\text{coll}})&=&2\eta_s\left[\frac{V_F}{E_{\text{coll}}}\asin\left(\sqrt{\frac{E_{\text{coll}}}{V_F}}\right)-\sqrt{\frac{V_F}{E_{\text{coll}}}-1}\right],
\label{eq1-4-1-5}
\end{eqnarray}
where $E_{\text{coll}}$ is the UCN kinetic energy at the point of collision. Note that each material has a unique value for the dimensionless number, $\eta_s$, associated with it.

The neutron loss channel in this apparatus is dominated by scattering off wall surfaces. In order to isolate the loss channel due to neutron decay, the method presented here involves fitting a well characterized, energy dependent, wall-scattering loss model to the number of stored UCNs in the chamber, as a function of storage time, in order to estimate the neutron lifetime. This method is functionally similar to the work presented in Refs.~\cite{Mohanmurthy2019-ju}, where a similar loss model was employed to extract the mean time between consecutive collisions of stored UCNs to be used in Ref.~\cite{Abel2021-hp}. In Refs.~\cite{Mohanmurthy2019-ju}, $\eta_s$ was allowed to be a free parameter, and the value of neutron lifetime was fixed to $\tau_n=879.4~$s. In this work, we fixed the values of $\eta_S$ to the values in literature, and allowed $\tau_n$ to be a free parameter. In Refs.~\cite{Mohanmurthy2019-ju}, a common energy independent loss coefficient $\eta_S$ was used. Here, care was taken to associate each surface with its specific value of $\eta_S$ and Fermi potential, $V_F$, as during storage the UCNs may scatter over a layer of dPS, coating the circular inner surfaces of the storage cylinder, or scatter over a layer of DLC, coating the flat ends of the storage cylinder. The values of $\eta_S$ and $V_F$ used are listed in Table~\ref{tab9-4}.

\begin{table}[h!]
\centering
\caption[]{Important surface parameters used in the loss model. The circular inner surface of the cylindrical UCN storage chamber is coated with dPS, and the flat inner surfaces are coated with DLC.}
\label{tab9-4}
\begin{tabular}{c | c c c c}
\hline
\hline
\diagbox{Surface}{Param.} & $V_F~\text{(neV)}$ && $\eta_S$ &\\
\hline
DLC & $(220\pm10)$ & \cite{Atchison2007-uz} & $(3.1\pm0.9)\times10^{-4}$ & \cite{Atchison2006-th}\\
dPS & $(165\pm10)$ & \cite{Bondar2017-fn} & $(3.0\pm1.0)\times10^{-4}$ & \cite{Bodek2008-ag}\\
\hline
\hline
\end{tabular}
\end{table}

\subsection{Characterization of the Energy Spectra}

A loss model based on scattering off walls demands characterization of the energy spectra of the UCNs. We use a generic 4-parameter energy distribution, as a function of energy of UCNs at the bottom of the storage chamber, $E(h=0)$, and at the beginning of storage phase (at $t_s=0$) using
\begin{eqnarray}
\mathcal{P}(E)=\mathcal{P}_0 \frac{E^{\rho}}{1+\exp{\frac{E-E_p}{w}}},
\label{eq9-5-5-7-1}
\end{eqnarray}
where $\mathcal{P}_0$ is a scaling constant, $\rho$ is the exponent of the leading edge of the distribution, $E_p$ is an upper energy cut-off value, and $w$ is a smearing parameter for the energy cut-off value. This distribution is additionally truncated at the lowest Fermi surface potential that the stored UCNs see, which for dPS is $V_F=165~$neV~\cite{Bondar2017-fn,Bodek2008-ag}, since the UCNs with energy above the Fermi surface potential of dPS are lost during storage. The parameterization shown in Eq.~\ref{eq9-5-5-7-1} is a modified version of that used in Ref.~\cite{Afach2015-ad}. The present approach, however, treats the exponent, $\rho$, as a free parameter, and considers an energy range beginning from zero. Further details about the model can be found in Refs.~\cite{Afach2015-ad,Mohanmurthy2019-ju}.

In order to study the time evolution of the energy spectrum we consider the following quantities \cite{Golub1991-ap}: (i) mean free path, given by $\lambda=4\cdot\text{volume}/\text{surface-area}$, (ii) bounce rate, given by $\nu(E_{\text{coll}})=(1/\lambda)\sqrt{(2E_{\text{coll}})/m_n}$, where $E_{\text{coll}} \approx E_{\text{bottom}} - m_n g\langle h\rangle$ is the collision kinetic energy at a point of height, $\langle h \rangle$, and (iii) loss probability per bounce, given by Eq.~\ref{eq1-4-1-5}.

\sloppy The numerical model of the storage curve consists of 5 free parameters: $\{\mathcal{P}_0,~\rho,~E_p,~w;~\tau_n\}$. The UCN loss rate, given by $1/\tau_{\mu} = \nu(E_{\text{coll}}) \cdot \mu(E_{\text{coll}})$, is used to calculate the attenuation at a given storage time for a certain energy. The energy spectrum at a time $t_s > 0$ is calculated by evolving the spectra at time $t_s = 0$ according to
\begin{eqnarray}
\mathcal{P}(E,t_s) = \mathcal{P}(E) \exp\left(-t_s \cdot \nu(E) \cdot \mu(E)\right).\label{eq9-5-5-7-1-b}
\end{eqnarray}
Finally, Eq.~\ref{eq9-5-5-7-1-b} is integrated over energy, $E$, in order to calculate the decay function which one would measure in an experiment. The data points of the resulting model storage curve are denoted by $n'(t_s)$, and can be expressed as:
\begin{eqnarray}
n'(t_s) = \int~dE \cdot \mathcal{P}(E,t_s).\label{eq9-5-5-7-1-c}
\end{eqnarray}
The goal of constructing the model decay curve, as described in Eq.~\ref{eq9-5-5-7-1-c}, was to find the distribution of the free parameters that best reproduces the measured decay curve shown in Figure~\ref{fig9-3} (Top). In Figure~\ref{fig9-3} (Top), the numerical model value for the decay curve, at the storage times where the experimental data points were also available, obtained by using the above formalism have also been plotted. Note that these numerical model values plotted in Figure~\ref{fig9-3} (Top) are for a single set of best fit parameters with the lowest reduced $\chi^2$. However, a wide range of parameter sets which also fit the decay curve data points within an acceptable reduced $\chi^2$ could be sampled. The range of residuals resulting from the parameter sets that passed both the Fisher statistical test (to be discussed in this section) and the center-of-mass constraint (to be discussed in the next section) have been shown in Figure~\ref{fig9-3} (Bottom).

\begin{figure}[h]
\includegraphics[width=\columnwidth]{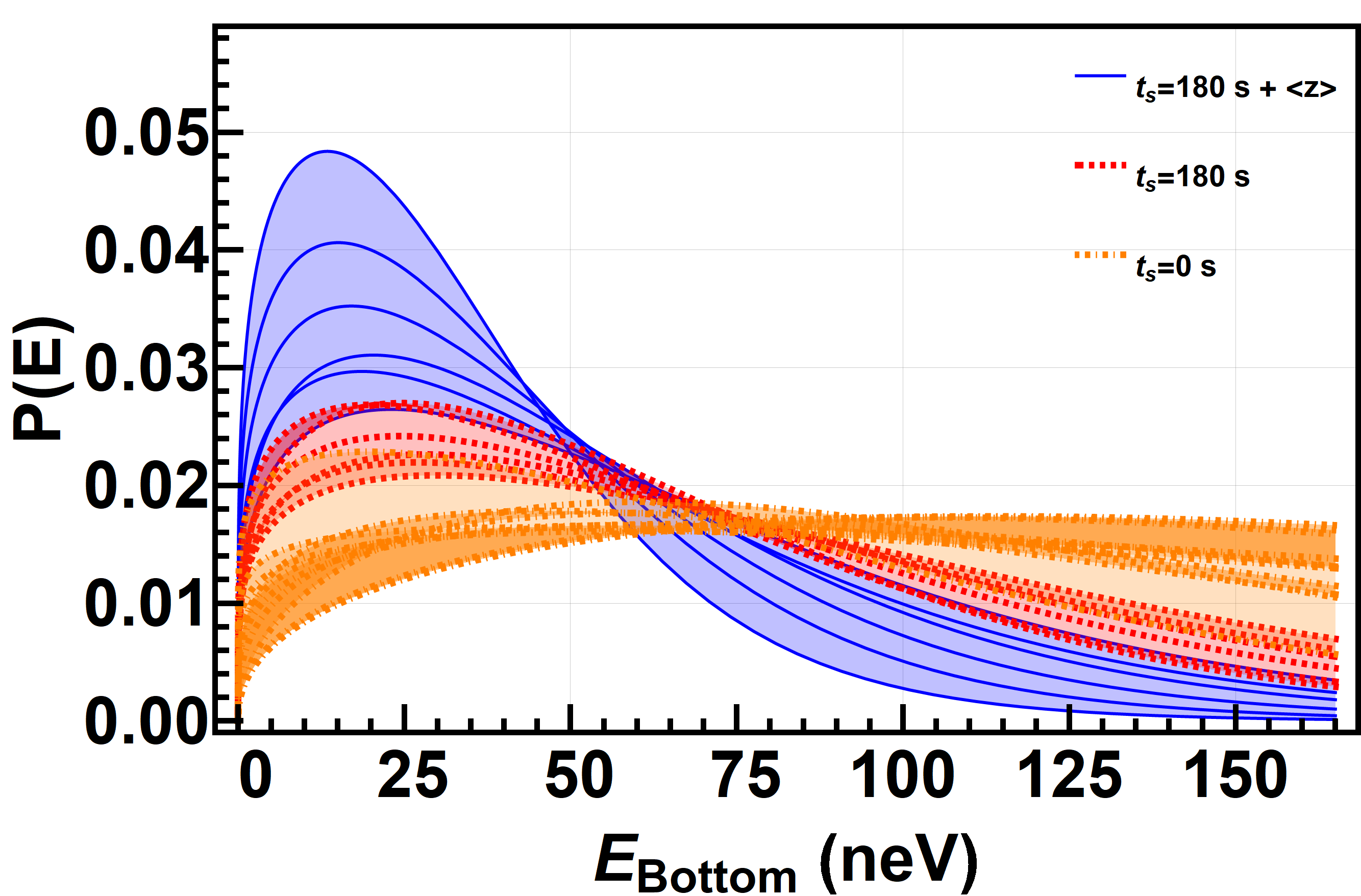}
\caption[]{Plot showing sample sets of energy distributions for the UCNs. The red curves represent those which used parameters of $\{\mathcal{P}_0,\rho,E_p,w;\tau_n\}$ that only passed the Fisher statistical test in Eq.~\ref{eq9-5-5-7-4} within a 68.3\% C.L. The curves in orange represent the spectra at the beginning of storage, $t_s=0$. The blue curves indicate those that, in addition to the Fisher test, also pass the center-of-mass constraint from Eq.~\ref{eq9-13b} at $t_s=180\,$s of storage time.}
\label{fig9-4}
\end{figure}

We used Monte-Carlo (MC) sampling to determine the confidence intervals of the sampled parameters in the model. The $5$ free parameters, $\{\mathcal{P}_0,~\rho,~E_p,~w;~\tau_n\}$, were randomly and uniformly sampled a large number of times to generate a set of initial energy spectra using Eq.~\ref{eq9-5-5-7-1}. In using the parameters for the Fermi potential of the surfaces, $V_F$, and the energy independent loss per bounce parameter, $\eta_S$, we choose to consider $99\% $ of the distributions in Table~\ref{tab9-4}. We sampled values of the neutron lifetime, $\tau_n$ from a range of up to nearly twice the neutron lifetime in literature, \emph{i.e.} in the range between $(50,1800)~$s, which is adequate to see the characteristic features of the model. The other parameters of $E_p$ and $w$ were sampled from a range of values between $(0,250)~$ neV, since these parameters do not constrain the UCN storage curve at larger values, near the upper limit of the sampled range. The number of initial fast UCNs diminishes over storage time and therefore have a negligible contribution at larger storage times for which the data was collected. Therefore, choosing the upper limit of sampling of parameters $E_p$ and $w$ well above the Fermi potential of the surfaces is sufficiently conservative. For a perfect Maxwell distribution, the value of the exponent is $\rho=1/2$ \cite{Golub1991-ap}, but it is relaxed to be a free parameter around the ideal value to accommodate spectra that are not strictly Maxwellian \cite{Zsigmond2018-pt}. Lastly, the parameter of $\mathcal{P}_0$ is simply a normalization that does not change the shape of the spectrum.

Comparing the computed storage curves (obtained for different sets of free parameters) with the measured storage curve in Figure~\ref{fig9-3} (Top) (from Ref.~\cite{Abel2019-rk}) allows us to compute a reduced $\chi^2$ for each sampled set of the $5$ parameters. In order to select sets of $5$ free parameters based on associated $\chi^2$, we employed the Fisher statistical test~\cite{James2006-qm,Particle_Data_Group2020-an}. The Fisher statistical test requires
\begin{eqnarray}
\frac{\chi^2}{\chi^2_{\text{min}}} \le 1+ \frac{\nu_1}{\nu_2-\nu_1}\mathcal{F}^{\alpha}_{\nu_1,(\nu_2-\nu_1)},
\label{eq9-5-5-7-4}
\end{eqnarray}
where $\mathcal{F}^{\alpha}_{\nu_1,(\nu_2-\nu_1)}$ is the Fisher function, $\nu_1=5$ is the number of model parameters, $\nu_2=34$ is the number of data points shown in Figure~\ref{fig9-3} (Top) (from Ref.~\cite{Abel2019-rk}), and $\chi^2_{\text{min}}$ is the minimum $\chi^2$ obtained from all the iterations of the MC sampling of the free parameters. For the full sample from the accepted parameter space, we considered a confidence range of up to 99.9\% C.L., by setting $\alpha = (1 - 0.999)$. For a $68.3\%$ C.L. range, one can set $\alpha$ accordingly. Only the samples of the $5$ free parameters that satisfy the Fisher statistical test are retained. The process of selecting the sampled spectra is similar to the process described in Refs.~\cite{Mohanmurthy2019-ju}. Example spectra that were selected using the above process can be found in Figure~\ref{fig9-4}.

\subsection{Constraints upon UCN center-of-mass offset}

Moments of the density distribution for the stored UCNs are key characterizing features of the ensemble of stored UCNs. Vertical striation of UCNs stored in a cylindrical chamber has been well characterized \cite{Afach2015-ad,Afach2015-en}. The loss model used in this work, and the associated energy spectra of UCNs, can also be used. The average height of the UCNs was calculated using the formula~\cite{Harris2014-kk,Pendlebury1994-nz}
\begin{eqnarray}
 \left<z\right>~=~\frac{E}{k}\left(k - 0.6 + 0.6\left(1-k\right)^{5/3}\right)-\frac{H}{2},
\label{eq9-5-5-7-5a}
\end{eqnarray}
where $k = 1$ when $E < mg(H=12~\text{cm})$, and $k = 1 - (1 -mgH/E)^{3/2}$ otherwise. For stable solutions in the parameter space, \emph{i.e.} excluding the sparsely populated regions, one obtains for the $68.3\%$ C.L
\begin{equation}
\left< z \right> = - (0.45 \pm 0.18)~\text{cm}. \label{eq9-13}
\end{equation}
The associated center-of-mass, $\left<z\right>$ (defined \emph{w.r.t.} to the geometric center of the storage chamber) have been plotted against the associated reduced $\chi^2$ (limited to $68.3\%$ C.L.) in Figure~\ref{fig9-5}.

The center-of-mass, $\left<z\right>$, was independently (and more precisely) extracted in the nEDM analysis~\cite{Abel2020-jr}. Even though the nEDM experiment dealt with polarized UCNs, and this effort deals with unpolarized UCNs, it has been previously shown that the energy spectra of the UCNs do not significantly vary between polarized and unpolarized species \cite{Abel2021-hp,Mohanmurthy2019-ju}. In order to improve our constraint on the energy spectrum, and thereby also on neutron lifetime, the value of \cite{Abel2020-jr}
\begin{equation}
\left<z\right>_{\text{PSI-nEDM}}=-(0.39\pm0.03)~\text{cm} \label{eq9-13b}
\end{equation}
from the existing nEDM analysis was used to further narrow the acceptance of energy spectra and sampled spectral parameters.

\begin{figure}[h]
\includegraphics[width=\columnwidth]{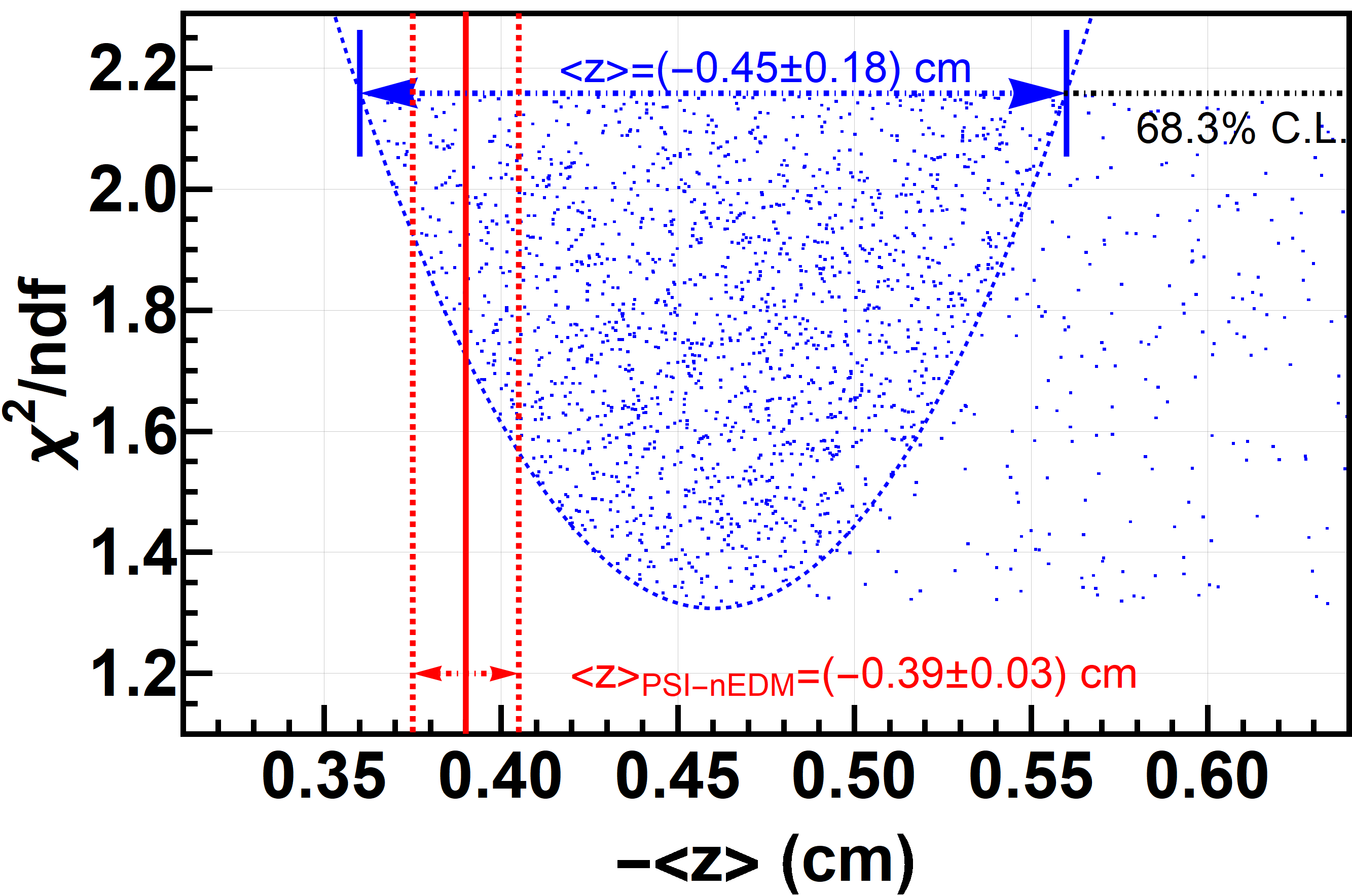}
\caption[]{Plot showing center-of-mass, $\langle z \rangle$, values along with their associated reduced $\chi^2$, for every set of five parameters sampled that passed the Fisher statistical test. A $68.3\%$ C.L. confidence interval was extracted from the numerical loss model alone, and has been indicated here graphically as as well as in Eq.~\ref{eq9-13}. The red vertical lines indicate the additional condition in Eq.~\ref{eq9-13b} used to further narrow down the accepted parameter space.}\label{fig9-5}
\end{figure}

When selecting for families of energy spectra, we also calculated the center-of-mass for the corresponding sampled parameters, and the sampled parameters and the corresponding energy spectra were retained only when the center-of-mass fell within the region indicated by PSI-nEDM in Ref.~\cite{Abel2020-jr}. For example, a set of sample spectra, calculated for $t_s=180\,$s of storage, which satisfy only the Fisher statistical test (with 68.3 \% C.L.) are shown in Figure~\ref{fig9-4} as red curves.  A second set of example spectra using the additional constraint from the center-of-mass is shown in the same figure by blue curves. The red curves only satisfy the Fisher statistical test, whereas the family of blue curves also satisfy the center-of-mass constraint in addition to satisfying the Fisher statistical test. Similarly, the range of residuals plotted in Figure~\ref{fig9-3} (Bottom) was obtained after applying the Fisher statistical test as well as the constraint from the center-of-mass offset. We have also applied the same constraint from the center-of-mass offset along with a Fisher statistical test when treating the neutron lifetime in the next section.

\subsection{Neutron lifetime extraction}

Neutron lifetime is one of the five sampled parameters in the neutron storage loss model described in previous sections. A reduced $\chi^2$ was associated with each set of $5$ sampled parameters. Figure~\ref{fig9-5} shows the reduced $\chi^2$ associated with each sampled value of the neutron lifetime. Each unique value for the neutron lifetime may have multiple values of reduced $\chi^2$ associated with it owing to the other $4$ sampled parameters in the model.  Selecting for the set of $5$ parameters which result in a good overall reduced $\chi^2$ using the Fisher statistical test narrows the samples down, and are further narrowed down by testing for the center-of-mass condition coming from the nEDM analysis. Ultimately, the best reduced $\chi^2$ for each value of neutron lifetime can be studied to obtain the values of the neutron lifetime that are consistent with the storage data shown in Figure~\ref{fig9-3} (Top).

\begin{figure}[h]
\includegraphics[width=\columnwidth]{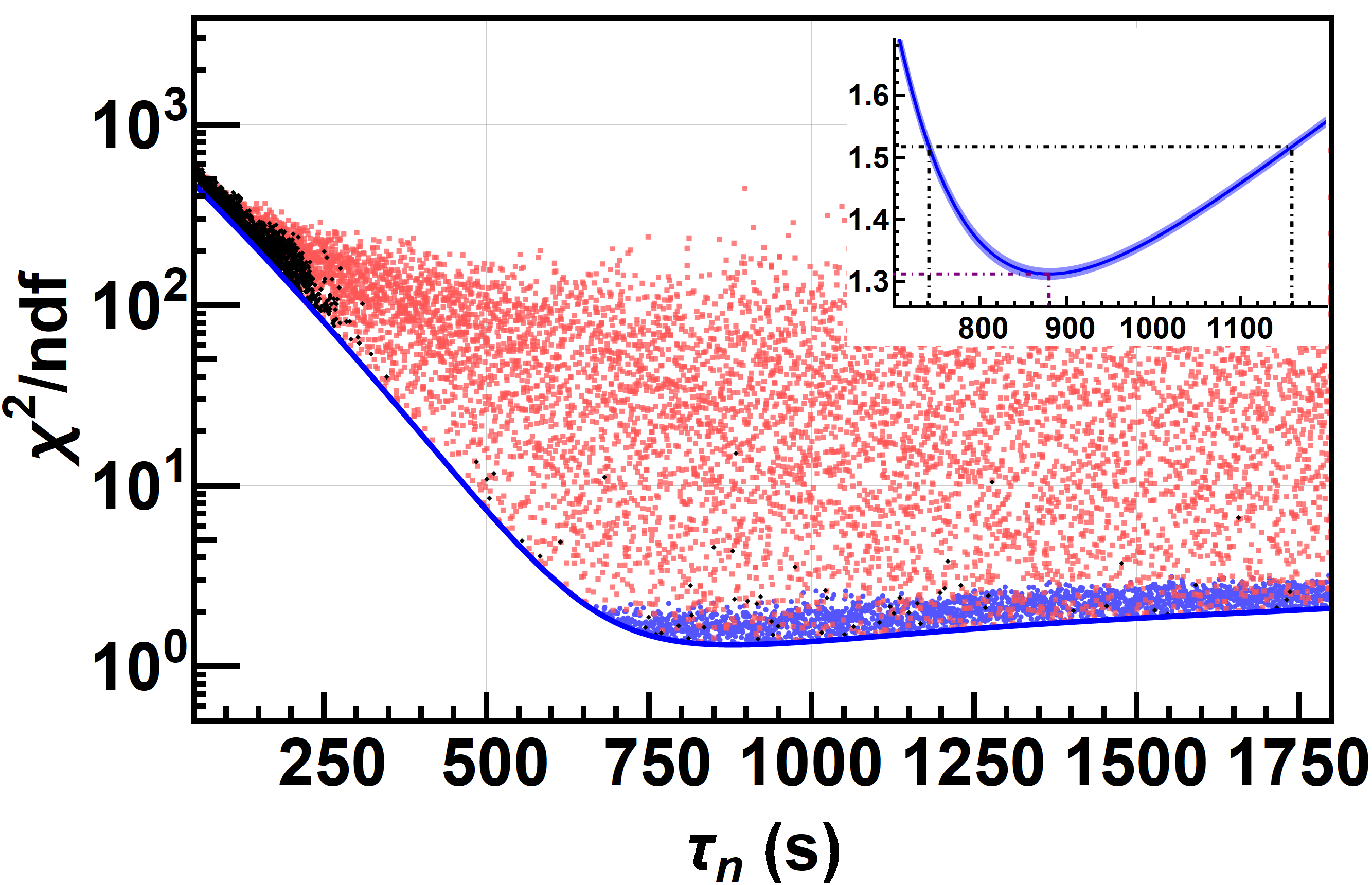}
\caption[]{Plot showing the values of sampled $\tau_n$ and the corresponding value of reduced $\chi^2$. Since there are four other freely sampled parameters of $\{\mathcal{P}_0,~\rho,~E_p,~w\}$ in the model, each value of $\tau_n$ may have multiple values of reduced $\chi^2$ associated with it. The red points represent all the sampled points, the black points represent samples which only passed the Fisher statistical test in Eq.~\ref{eq9-5-5-7-4}, while the blue points also satisfied the center-of-mass constraint in Eq.~\ref{eq9-13b}. The inset shows the best reduced $\chi^2$ as a function of sampled values of $\tau_n$, where the blue region is the $68.3\%$ C.I. uncertainty obtained from repeating the MC sampling.}
\label{fig9-5}
\end{figure}

Once the loss channel due to scattering off the walls has been well characterized, the next most dominant loss channel is due to the neutron $\beta$~decay. Consequently, after accounting for the neutrons lost to scattering off the walls, the reduced $\chi^2$ has a minima around the neutron lifetime. At choices of low lifetime values, the neutrons disappear at a higher rate, and for choices of high lifetime values, the neutrons survive for longer time than the data indicates, both leading to higher reduced $\chi^2$ values.

In order to obtain high density sampling, the Fisher test was applied to sets of $5$ sampled parameters where the lifetime values were the same, and samples which passed the Fisher test have been marked in black in Figure~\ref{fig9-5}. These values need not necessarily satisfy the center-of-mass condition coming from the nEDM analysis. The samples which also satisfy the center-of-mass condition coming from the nEDM analysis have been indicated in blue in Figure~\ref{fig9-5}. Other points indicated in red correspond to sets of each of the $5$ sampled parameters which neither passed the statistical test nor the center-of-mass condition. The Fisher test was further used to compare the global minimum reduced $\chi^2$ with the minimum reduced $\chi^2$ for each unique value of sampled lifetime to obtain the confidence interval
\begin{equation}
\tau^*_n=879^{+279}_{-138}~\text{s~~(68.3\% C.L.)}. \label{eq9-22} 
\end{equation}
The global minimum reduced $\chi^2$ was $\chi^2/ndf=1.31$ obtained at the sampling value of $\tau_n=879.7~$s. This best reduced $\chi^2$ is consistent with the values reported from comparable models in Refs.~\cite{Afach2015-ad,Mohanmurthy2019-ju}. Note that the $68.3\%$ confidence interval shown in Eq.~\ref{eq9-22} has also been indicated in Figure~\ref{fig9-5}.

The above numerical MC sampling was repeated over $250~$times. The standard deviation of the minumum reduced $\chi^2$ for each unique value of lifetime, coming from repetition of the process, is shown as shaded region in the inset of Figure~\ref{fig9-5}, and the mean is shown as the solid blue line. The uncertainty from repeatability impacted the result in Eq.~\ref{eq9-22} by less than $1\%$, showing that the sample size considered is sufficiently large enough and the error related to the numerical sampling is negligible.

\section{Discussion of the Sources of Errors}

In this section, we will discuss the sources of the uncertainty contributing to the result shown in Eq.~\ref{eq9-22}. The input parameters into the model, whose uncertainties are numerically translated to the uncertainty in the lifetime, are the three parameters: (i) energy independent loss per bounce, $\eta_S$, (ii) the normalized neutron counts in Figure~\ref{fig9-2} (Top), $n_i$, and (iii) the Fermi potential of the surfaces, $V_F$. The uncertainties associated with the values of $\eta_S$ and $V_F$ used in the model maybe found in Table~\ref{tab9-4}, and the uncertainties associated with the normalized neutron counts come from propagating the counting statistics (Poisson errors) of the emptying and monitor counts. In order to understand the propagation of these sources of uncertainty to the uncertainty in the value of neutron lifetime, all but one uncertainty was set to zero, and the process was run as usual. This is the same process that was employed in Ref. ~\cite{Abel2021-hp} to understand the contribution of the individual parameter's uncertainty to the uncertainty on the final value. The resulting uncertainties are presented in Table~\ref{tab9-8}.

\begin{table}[h!]
\centering
\caption[]{Table showing the uncertainty contributions of the parameters of $V_F$, $n$, and $\eta_S$, to the value of $\tau^*_n$ extracted in Eq.~\ref{eq9-22}.}
\label{tab9-8}
\begin{tabular}{c | c c}
\hline
\hline
{\small Errors from} & $+\sigma_{\tau^*_n}~$(s) & $-\sigma_{\tau^*_n}~$(s)\\
\hline
$V_F+n+\eta_S$ & $279$ & $138$\\
\hline
$V_F$ & $42$ & $21$\\
$n$ & $158$ & $78$\\
$\eta_S$ & $226$ & $112$\\
\hline
\hline
\end{tabular}
\end{table}

Since the uncertainties are asymmetric around the central value, Table~\ref{tab9-8} reports two values, each corresponding to one side of the central value. It is clear that the largest source of uncertainty is the value of energy independent loss per bounce, $\eta_S$, followed by the uncertainty coming from the normalized neutron counts, $n$, and lastly, from the value for the Fermi potential of the surfaces, $V_F$. While it is possible to obtain a factor of $2-3$ improvement in the uncertainty associated with the neutron lifetime extracted by improving the uncertainties in the input parameters of $V_F$, $n$, and $\eta_S$, by an order of magnitude, it is still not feasible to bring the final uncertainty on the neutron lifetime extracted down to few seconds using this data set.

The other sources of uncertainty that have not been shown in Table~\ref{tab9-8} are the uncertainties coming from (i) the length of time for which the storage curve data was collected, (ii) the interval in which the data was collected, and (iii) the uncertainty associated with $\langle z \rangle$. Given the capacity of the storage chamber in Ref.~\cite{Abel2019-rk}  coupled with the neutron density of the source, it is hard to extrapolate statistically significant number of neutrons beyond about $1000\,$s. Fitting the loss model using storage data that were at least twice as long as the final neutron lifetime, $[50,~950]\,$s, was essential to capture the effects of neutron $\beta$-decay using this model. Furthermore, in order to improve the uncertainty on the final neutron lifetime extracted, it is vital to capture the effects of neuron loss early in the storage phase by collecting storage curve data with a frequency of every few seconds. When the uncertainties in the values for $V_F$, $n$, and $\eta_S$ are improved, narrowing down the acceptable sample space using a cut applied from the value of $\langle z \rangle$ from a different, more precise analysis, in order to improve the uncertainty in the neutron lifetime extracted, may not be required. The uncertainties shown in Table~\ref{tab9-8} already incorporate the uncertainties from these three sources.

It is conceivable that the minimum interval with which the data was reported in Ref.~\cite{Abel2019-rk} was limited by the time it takes to fill and close the storage chamber by moving the mechanical UCN switch and the UCN shutter below the storage chamber. Neutrons may decay while they are filled or emptied out of the storage chamber, so a time period of $2\tau_{\text{emp}}\approx(22.6\pm0.4)~$s \cite{Abel2019-rk} has been added to the time interval defined in the data acquisition system, in order to obtain an effective storage time. Here $\tau_{\text{emp}}$ is the filling or emptying time constant. The interval of collecting storage curve data is limited by the uncertainty associated with the filling or emptying time constant.

Taking the above effects into consideration and using a toy data set for the storage curve, in order to obtain a measurement of the neutron lifetime using the above method to a precision of $1\%$, storage curve data would need to be collected for at least three times the neutron lifetime with a frequency of every second. At this level of precision, loss channels due to absorption and imperfect storage due to leakage would need to be equally well characterized. In addition, the sources of uncertainties coming from the input parameters of $V_F$, $n$, and $\eta_S$, would need to be improved by nearly $2$ orders of magnitude. In terms of the neutron count, this implies that the number of neutrons counted after storage needs to be improved by nearly $4$ orders of magnitude, owing to the counting statistics that dictates the associated uncertainty.

\section{Conclusion}

The statistical uncertainty portion of the neutron lifetime in Eq.~\ref{eq9-22} can be attributed to the uncertainty arising from the neutron counting statistics of the normalized neutron counts in Figure~\ref{eq9-2}, indicated in Table~\ref{tab9-8}. The systematic uncertainty portion of the neutron lifetime can be attributed to the uncertainty arising from the combination of uncertainties in the parameters $\eta_S$ and $V_F$. The result in Eq.~\ref{eq9-22} can then be written as a sum of statistical and systematic errors as
\begin{equation}
\tau^*_n=879~({+158}/{-78})_{\text{stat.}}~(+230/-114)_{\text{sys.}}~\text{s~~(68.3\% C.L.)}. \label{eq9-23}
\end{equation}
Clearly the result is dominated by the systematic portion of the uncertainties. This result may also be interpreted in terms of an interval at the $95\%$ C.L. by
\begin{equation}
\tau^*_n \in (688,1400)~\text{s~~(95\% C.L.)}. \label{eq9-23}
\end{equation}

Other MC simulations have been performed, particularly in Refs.~\cite{Serebrov2009-dv,Serebrov2010-ij}. Ref.~\cite{Serebrov2009-dv} corrects the value in Ref.~\cite{Mampe1989-ih} by around $-6\,$s by including quasielastic scattering of UCNs over a liquid wall, and Ref.~\cite{Serebrov2010-ij} corrects the value of the lifetime in Ref.~\cite{Arzumanov2000-tm} by a similar amount. These MC simulations use the same technique used by the corresponding experiments, where the loss factor associated with UCNs bouncing off walls of the storage chamber is accounted for by extrapolating the inverse of the mean free path to zero.

These simulations also characterize the energy spectra of the UCNs as a function of storage time, but fail to incorporate the uncertainty on these energy spectra. In our simulation, described in this paper, no such extrapolation is used, and the storage curve is directly (numerically) fitted to the loss model to characterize the loss channel coming from UCNs bouncing off the walls of the storage chamber. The uncertainties from the parameters $\eta_s$ and $V_F$, besides the uncertainties from the other two degrees of freedom in our loss model, $E_p$ and $w$, taken together contribute to a large uncertainty in the energy spectra of UCNs as a function of storage time. Such an uncertainty in the energy spectra leads to a large possible range for the acceptable number of times UCNs may have bounced off the walls during storage. A large uncertainty in the energy spectra makes characterizing the loss channel due to UCNs scattering off the wall surfaces harder, and makes the associated uncertainties larger. A large uncertainty in the energy spectra also implies that a larger spread of neutron lifetime may be consistent with the storage curve data. We therefore indicate that the previous MC simulations of UCN storage experiments underestimate their errors.

This effort uses data in Ref.~\cite{Abel2019-rk}, that was itself reported to have been collected using the nEDM apparatus. It was not designed or optimized to measure the neutron lifetime. Just like the early UCN storage experiments, this effort was dominated by losses due to neutron scattering off the walls of the storage chamber. But, we were for the first time able to develop a full fledged loss model that includes channels of loss due to scattering off the walls, due to imperfect storage of UCNs in the storage chamber, and of course due to neutron $\beta$~decay. The primary improvement presented here is to account for and propagate the sources of uncertainty upon the characterization of the family of energy spectra for the UCNs.


\funding{One of the authors, P. M., would like to acknowledge support from the SERI-FCS award \# 2015.0594, Sigma Xi grants \# G2017100190747806 and \# G2019100190747806, and the Ivy Plus Exchange program.}

\acknowledgments{We would like to acknowledge the grid computing resource provided by the Research Computing Center at the University of Chicago \cite{[rcc]}.}

\end{paracol}
\reftitle{References}

%

\end{document}